\documentstyle[aps,multicol,psfig]{revtex}

\newcommand{\PRB}[1]{Phys.\ Rev.\ B {\bf #1}}
\newcommand{\PRE}[1]{Phys.\ Rev.\ E {\bf #1}}
\newcommand{\PRL}[1]{Phys.\ Rev.\ Lett.\ {\bf #1}}
\newcommand{\JAP}[1]{J.\ Appl.\ Phys.\ {\bf #1}}
\newcommand{\PLA}[1]{Phys.\ Lett.\ A {\bf #1}}
\newcommand{\JPA}[1]{J.\ Phys.\ A {\bf #1}}
\newcommand{\PD}[1]{Physica D {\bf #1}}

\begin{document}


\draft

\title{Scaling in the time-dependent failure of a fiber bundle
with local load sharing}

\author{Shu-dong Zhang\cite{em1}}
\address{Institute of Low Energy Nuclear Physics,
Beijing Normal University, Beijing 100875, China\\
CCAST(World Laboratory), P.O.Box 8730, Beijing 100080, China\\
Beijing Radiation Center, Beijing 100088\\
Institute of Theoretical Physics, Beijing Normal University, Beijing 100875\\
}


\date{\today}

\maketitle

\begin{abstract}
We study the scaling behaviors of a time-dependent fiber-bundle model
with local load sharing.
Upon approaching the complete failure of the bundle, the breaking rate
of fibers diverges according to $r(t)\propto (T_f-t)^{-\xi}$,
where $T_f$ is the lifetime of the bundle, and
$\xi \approx 1.0$ is a quite universal scaling exponent.
The average lifetime of the bundle $<T_f>$ scales with the system size as
$N^{-\delta}$, where $\delta$ depends on the distribution of individual fiber
as well as the breakdown rule.
\end{abstract}


\maketitle

\begin{multicols}{2}
\narrowtext

\section{Introduction}
The failure of disordered materials under load is a complicated phenomenon,
the modelling of which is a subject of great interest because it forms the
basis of numerous applications  from space technology to paper
making~\cite{Hermann&Roux1990}. The failure process also represents an
important class of pattern formation and scaling
problems~\cite{Newmanetalphysicad1994}.
The fiber-bundle model, as a simple and interesting theoretical model in this 
field, has been studied extensively. 
The early studies on the static fiber-bundle model 
might be traced back to the work by Daniels~\cite{Daniels1945},
while the time-dependent method to the model was proposed
by Coleman~\cite{Colemanjap1958}. In a recent paper~\cite{Gomezetalpre1998},
Gomez {\em et al} developed a probabilistic method for solving
the time-dependent model.
In the static model, each fiber in the bundle is
assumed to have a strength threshold, a load above which will 
break it instantly, while a load below which does no harm.
In the time-dependent model, each fiber is assumed to have a lifetime under
a given load history, and it breaks because of fatigue.
The load-sharing rules, which describe how the load of a broken element is
transferred to survival elements, are essential to the definition of the model. 
In what is called Equal Load Sharing (ELS) model, the total load of the bundle 
is equally shared by all survival fibers, 
while in the Local Load Sharing (LLS) model 
the load of a broken fiber is transferred to its nearest neighbors. 
Hierarchically organized fiber bundle was also
proposed, and has received much attention especially in the geophysical
literature~\cite{Newmanetalpre1995,Turcotte1997}.
Various aspects of the fiber-bundle model have been investigated,
such as the strength distribution for static
model~\cite{Daniels1945,Harlow&Phoenix1978ab} and the lifetime distribution
for dynamic one~\cite{Colemanjap1958,Phoniex&Tierney1983}.
In this paper, we will study an LLS time-dependent model, and
investigate the scaling behaviors in its breaking process.

Let us consider a fiber bundle consisting of $N$ fibers.
We assume that when a fiber is subjected to a load history $\sigma (t)$,
some damage will accumulate, which is described by
\begin{equation}
d(t) = \int_{0}^{t}\nu[\sigma (\tau)]d\,\tau ,
\label{d(t)}
\end{equation}
where the load-dependent $\nu (\sigma)$ is
introduced as a hazard rate, which is usually referred to as 
{\em breakdown rule}~\cite{Colemanjap1958} in the literature.

A fiber, say fiber $i$, is assumed to have an endurance threshold 
(or say, critical damage) $d_i^c$, which is drawn from a
cumulative distribution
\begin{equation}
P(d_i^c<d)=1-\exp \left[ -\Psi (d)\right],
\label{P(d^c)}
\end{equation}
where $\Psi (x)$ is the {\em shape function}. Previous theoretical and
experimental work~\cite{Colemanjap1958,Phoniex&Tierney1983} favors a shape
function of the form
\begin{equation}
\Psi (x)=x^{\beta}.
\label{Psi(x)}
\end{equation}

As for the breakdown rule $\nu (\sigma)$, two special forms are widely used 
in the literature: the power-law form
\begin{equation}
\nu _p(\sigma)=\nu _0\left( {\sigma \over \sigma _0}\right)^{\rho},
\label{nu_p}
\end{equation}
and the exponential form
\begin{equation}
\nu _e(\sigma)=\phi _0\exp \left({\eta \sigma \over\sigma _0}\right),
\label{nu_e}
\end{equation}
with $\nu _0$, $\sigma _0$, $\rho$, $\phi _0$, $\eta$ all positive constants.

Under load each fiber will break when the damage accumulated exceeds its 
endurance threshold, and all fibers will break eventually, leading to
the  complete failure of the bundle.
Let us denote the total load on the bundle by $N\sigma$.  In general, 
$\sigma$ is a function of time. For example, it can be a linearly
increasing function or a periodic function of time~\cite{Colemanjap1958}.
In this paper, we will consider the simple case that $\sigma$ is a constant. 
In the following numerical calculations, if not otherwise specified, the
load is set to be $\sigma =\sigma_0$. It should be noted that although 
the total load on the bundle is constant, the loads on the individual fibers
$\sigma _i(t)$'s are not.

\section{The LLS model}

We consider a fiber-bundle model with the LLS rule.
$N$ fibers are arranged evenly on a circle, and each of them has
two adjacent neighbors.
The total load on the bundle $N\sigma$, kept constant in this study,
is shared by survival fibers. A survival fiber $i$ carries the load
$\sigma _i=K_i\sigma$, where the concentration factor
$K_i=1+(l_i+r_i)/2$. Here $l_i$ ( $r_i$ ) is the number of broken fibers
on the left (right) of fiber $i$.
It is clear that $\sum _{i}K_i=N$, so the total load is conserved. 
With such a load sharing rule, the load of a broken fiber is transferred to
the survival neighbors on both sides. Note that this rule is different from 
the one-side case~\cite{Gomezetalprl1993}, in which the load of a broken fiber 
is transferred only to its neighbor on one side.

This LLS fiber-bundle model was in early years developed by Harlow
and Phoenix\cite{Harlow&Phoenix1978ab} to
model the failure of a unidirectional composite material under tensile loads.
The model has ever since drawn much attention of many authors.  In recent 
years, the static LLS fiber-bundle model was studied  in terms of the burst-size
distribution~\cite{Hansen&Hemmerpla1994,Zhang&Dingjpa1995,Klosteretalpre1997}
and the failure probability of the bundle under a given
load~\cite{Leath&Duxburyprb1994,Zhang&Dingprb1996}.
In this study, we will focus on the scaling behaviors of 
this dynamic LLS fiber-bundle model.

\section{Scaling of Breaking Rate with Time to Failure}
Let $N_f(t)$ be the number of broken
fibers in the bundle at time $t$, with $N_f(0)=0$ and $N_f(T_f)=N$, where 
$T_f$ is the lifetime of the whole bundle.
The breaking rate of the bundle is defined as
\begin{equation}
r(t)={\delta N_f(t)\over \delta t}
\end{equation}

\begin{figure}
\centerline{\psfig{file=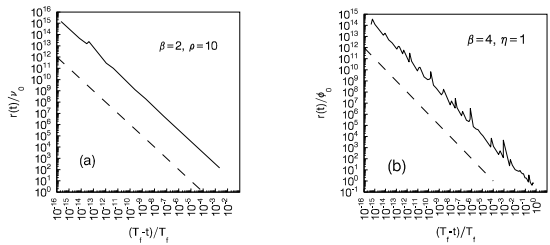,width=3in,angle=270}}
\caption{
The breaking rate $r(t)$, defined in the text, scales with
the time to failure as $(T_f-t)^{-\xi}$, where $\xi \approx 1.0$ is
a quite universal value. (a) Using the power-law breakdown rule
(\protect\ref{nu_p}) with $\rho=10$, and the shape function
(\protect\ref{Psi(x)}) with $\beta=2$. (b) Using the exponential
breakdown rule (\protect\ref{nu_e}) with $\eta=1.0$, and shape-function 
parameter $\beta=4$.
In both (a) and (b), the system sizes are $N=100$, and the dashed lines show
the curves for $y\propto x^{-1}$ for reference.
}
\label{figbreakingrate}
\end{figure}

We have performed extensive Monte Carlo simulations of the breaking process 
of the time-dependent fiber-bundle model with LLS, and found that in a wide 
range of parameter value the breaking rate $r(t)$, upon approaching the 
complete failure, scales with the time to failure as
\begin{equation}
r(t)\propto (T_f-t)^{-\xi}
\label{r(t)}
\end{equation}
and the scaling exponent $\xi\approx 1.0$ is of a quite universal value.
Examples of the behavior of the breaking rate are shown in
Fig.~\ref{figbreakingrate}. 
In this log-log plot, dashed lines with slope $-1$ are also shown for 
reference. The numerical
results are not very smooth because of fluctuation, but the general
trend of the breaking rate $r(t)$ agrees well with Eq.(\ref{r(t)}).

\begin{figure}
\centerline{\psfig{file=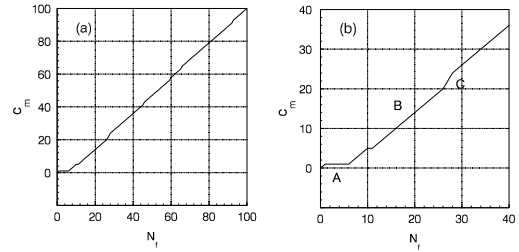,width=3in,angle=270}}
\caption{An example of the evolution of the biggest crack in the failure
process of the fiber bundle. The exponential breakdown rule is used with
$\eta=1$. The other parameters are $N=100$, $\beta=4$.
(b) is a part of (a) enlarged.
}
\label{figcrack}
\end{figure}

In what follows, we try to understand the scaling behavior (\ref{r(t)}) 
through analytical treatment. In the discussion, we take the limit 
$N\rightarrow \infty$.  Let us call the connective broken fibers bounded by 
unbroken ones as a {\em crack}. The {\em size } of a crack is the number of 
broken fibers. Because of the local load-sharing rule, the fibers bounding 
a larger crack experience heavier load than
those bounding smaller ones.  So when a major crack is
formed in the bundle, breaking will mostly occur along it. 
In other words, it is the fibers adjacent to the major crack that will 
most probably break in the next step. This can be seen from the evolution 
of the size $c_m$ of the biggest crack. 
Fig.~\ref{figcrack} shows $c_m$ versus the total number of 
broken fibers in the bundle. At the early stage of the
failure process, $c_m$ remains constant for some time ( $A$ ),
which indicates that small cracks nucleate at different locations.
As more and more fibers break, some small cracks will coalesce or grow to 
form a major crack, and then the major crack grows, which is reflected in 
this figure by a linear increase of $c_m$ with $N_f$ with slope $1$ ( $B$ ). 
During its growth, the major crack may also coalesce with some small cracks and 
become even larger, indicated in the figure by local slopes steeper than $1$ at 
some points ( e.g., $C$ ).

Suppose the size of the major crack is $N_f(t)-k$, where $k$ is the
number of failed fibers which do not belong to the major crack. The
loads on the fibers adjacent to the major crack are 
$[1+(N_f-k)/2]\sigma$,
so damage will accumulate in these fibers with the rate
$\nu ([1+(N_f-k)/2]\sigma)$. The breaking rate of these fibers can be
assumed to be proportional to $\nu(\cdot)$, and one has
\begin{equation}
r(t)={d\, N_f(t)\over d\,t}
=A(t)\nu \left( \left[1+\frac{N_f-k}{2}\right]\sigma\right),
\end{equation}
where $A(t)$ is a factor that depends on the accumulated damages in the fibers
and their endurance thresholds. An exact calculation of $A(t)$ is extremely
difficult and might be impossible. We assume that the variance of $A(t)$ 
is unimportant and take $A$ as a constant for simplicity. The validity of
this assumption is verified by the agreement with numerical results.
Note that sometimes a fiber adjacent to the major crack happen to be 
also adjacent to a small crack, resulting in a little more load on 
it, the influence of which on the breaking rate however, is 
negligible upon approaching the complete failure. 

For the exponential form of breakdown rule (\ref{nu_e}), we have
\begin{equation}
{d\, N_f(t)\over d\,t}=A\phi _0
\exp \left[\eta\left(1+{N_f-k\over 2}\right){\sigma \over \sigma _0}\right],
\end{equation}
and therefore,
\begin{equation}
r(t)=\alpha ^{-1}(T_f-t)^{-1},
\end{equation}
where $\alpha=\eta\sigma/(2\sigma _0)$,
$T_f$ is the value of time that gives $N_f(T_f)\rightarrow \infty$.

For the power-law form of breakdown rules (\ref{nu_p}), 
\begin{equation}
{d\, N_f(t)\over d\,t}=A\nu _0
\left[\left(1+{N_f-k\over 2}\right){\sigma \over \sigma _0}\right]^{\rho},
\end{equation}
and
\begin{equation}
r(t)=C\left[{\rho-1\over 2}C(T_f-t)\right]^{\rho\over 1-\rho}
\propto (T_f-t)^{-1-{1\over \rho -1}}
\end{equation}
with $C=A\nu _0(\sigma /\sigma _0)^{\rho}$, and $N_f(T_f)\rightarrow \infty$.
So $\xi =1+1/(\rho -1)$.  Since $\rho$ is of quite large value,
typically between $10$ and $80$~\cite{Phoniex&Tierney1983},
it is not surprising that $\xi \approx 1.0$ in the numerical simulations.

\section{Lifetime of the bundle}
\begin{figure}
\centerline{\psfig{file=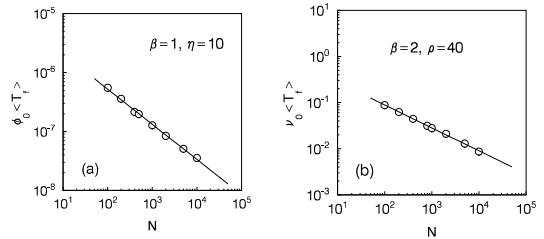,width=3in,angle=270}}
\caption{
The average lifetime of the bundle scales with system size
$N$ according to a power law. The circles are results from
numerical simulations with at least $100$ samples, the solid line is 
for the power-law fit $y=ax^{-\delta}$ to the numerical data.
(a) $\beta=1$, $\eta=10$, $a=8.14\times 10^{-6}$, $\delta=0.60$.
(b) $\beta=2$, $\rho=40$, $a=0.87$, $\delta=0.50$.
}
\label{figlifetime}
\end{figure}

In deducing the scaling of the breaking rate, we have taken the 
thermodynamic limit by setting $N_f(T_f)=\infty$.
In numerical simulations however, we cannot realize infinite system 
size. 
Given the local load-sharing rule, the lifetime $T_f$ of a fiber bundle depends 
on the endurance of each fiber.  
Due to fluctuation, $T_f$ is different from bundle to bundle. 
Since the fluctuation is related to the system size, the average lifetime 
$<T_f>$ of the bundle should in principle depend on $N$, which is known as 
size effect.
We found that in general the average life time $<T_f>$ scales with the system 
size as
\begin{equation}
<T_f> \propto N^{-\delta},
\label{N^delta}
\end{equation}
where $<\cdots >$ means the ensemble average.
Some of the numerical results are shown in Fig.~\ref{figlifetime},
in which the power-law fit to the data is quite good.
Some other forms of fit to the data were also tried, but none of them is better
than the power law.
It should be noted that in the static LLS model the average
strength of the bundle follows a logarithmic dependence on
the system size~\cite{Gomezetalprl1993,Zhang&Dingpla1994}.

The exponent $\delta$ for the power law, however, is not of a universal value. 
It depends on the breakdown rule as well
as the distribution of damage endurance for individual fiber. We performed 
extensive numerical simulations to explore the relation between the
exponent $\delta$ and the parameters $\beta$, $\rho$ and $\eta$.
Some results are listed in Table \ref{tabledelta}.
There seems no simple general expression relating
$\delta$ to $\beta$, $\rho$, and $\eta$. 
For some limiting cases, however, we can get a simple relation.
From Table \ref{tabledelta}, one can see that when $\rho$ or $\eta$ is large,
the value of the exponent $\delta$ is very close to $1/\beta$. This result
can be understood by the lifetime distribution of the fiber bundle.
When $\rho$ or $\eta$ is large, the fiber bundle
breaks in the following way: when the weakest fiber breaks, it will form the
crack that leads to the failure of the whole bundle. So the lifetime
of the bundle will be expended mostly in the weakest fiber,
and is thus determined by it.
From Eqs. (\ref{d(t)}), (\ref{P(d^c)}) and (\ref{Psi(x)}), 
the lifetime of an individual fiber under a constant load $\sigma$, is 
distributed as
\begin{equation}
P(t_f<t)=1-e^{-[\nu (\sigma)t]^{\beta}}.
\end{equation}

For a bundle of $N$ fibers, if the bundle's lifetime is determined by
the lifetime of its weakest element, the lifetime distribution for such 
a bundle is, by the weakest-link rule and when $N$ is large,
\begin{equation}
P(T_f<t)=1-e^{-N[\nu (\sigma)t]^{\beta}}.
\label{P(T_f<t)}
\end{equation}
And this is the Webull distribution, with which the average lifetime of the bundle is
\begin{equation}
<T_f>=\int_0^{\infty}t\,dP(T_f<t)
=\int_0^{\infty}t\,d(1-e^{-N[\nu (\sigma)t]^{\beta}}).
\end{equation}
Changing the variable of integration $Nt^{\beta}=\tau ^{\beta}$, one gets
\begin{equation}
<T_f>=N^{-1/\beta}\int_0^{\infty}\tau \,
d(1-e^{-[\nu (\sigma)\tau]^{\beta}}).
\end{equation}
The integration in the above equation is independent of $N$, 
so  $<T_f>\propto N^{-1/\beta}$, and $\delta=1/\beta$.

\begin{figure}
\centerline{\psfig{file=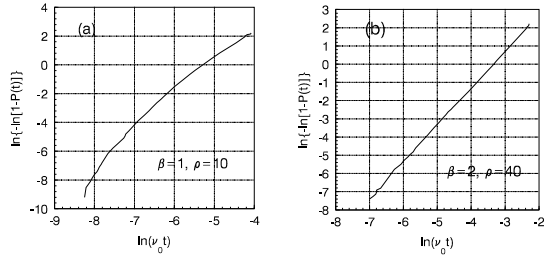,width=3in,angle=270}}
\caption{
The lifetime distribution of the LLS fiber bundle.
The results in this figure are from simulations of $10^4$ samples.
(a) $\beta=1$, $\rho=10$, $N=1000$. The curve is not a straight line.
(b) $\beta=2$, $\rho=40$, $N=800$. The curve is a quite straight line,
indicating a Webull distribution $P(t)=1-\exp (-at^m)$.
The best linear fit to the numerical data in (b) gives the
slope $m\approx 2.03$.
}
\label{figwebull}
\end{figure}

From the numerical results, we notice that $\delta=1/\beta$ is not satisfied
by all values of $\rho$ and $\eta$.
The deviation of $\delta$ from $1/\beta$ may indicate the 
deviation of the lifetime distribution from the Webull distribution.
In Fig.~\ref{figwebull}, we plot the lifetime distribution of the fiber bundle
with Webull axes, that is, to plot $\ln \{-\ln [1-P(t)]\}$ versus $\ln t$.
If the distribution is of Webull form, $P(t)=1-\exp (-at^m)$, one should see
a straight line in such a plot, and the slope of the line gives the Webull
modulus $m$. For the case $\beta=2$ and $\rho=40$ (Fig.~\ref{figwebull}.b),
we get a quite straight line, and the best linear fit to the distribution 
curve gives the Webull modulus $m\approx 2.03$, very close to $\beta=2$.
Notice that for this case $\delta \approx 0.50=1/\beta$.
For the case $\beta=1$ and $\rho=10$ (Fig.~\ref{figwebull}.a), however,
the distribution curve is not a straight line,
indicating that the lifetime is not very well Webull distributed.
For this case $\delta\approx 0.49$, which is quite different from $1/\beta=1.0$.

In the early studies on the lifetime distribution, Phoenix and his collaborators
~\cite{Phoniex&Tierney1983} were able to obtain an approximation to the lifetime
distribution of the fiber bundle, which was also of Webull form.
Their results were based on the idea that whenever a crack of
critical size, called $k^*$-crack in their paper, emerges in the system,
the bundle will fail instantly.

\section{Conclusions}

In conclusion, we have studied some scaling behaviors of the
time-dependent fiber-bundle
model with LLS rule. In a quite wide range of parameter value, 
the breaking rate scales with the time to failure as
$(T_f-t)^{-1}$. The average lifetime of the bundle scales with
system size as $N^{-\delta}$, with $\delta$ dependent on the
breakdown rule and endurance distribution of individual fiber.
In the limiting cases that $\rho$ or $\eta$ is very large, the lifetime
distribution of the bundle can be well approximated by a Webull form,
and the Webull modulus for this distribution is just the shape-function
parameter $\beta$, and the scaling exponent $\delta=1/\beta$.

\centerline{\bf ACKNOWLEDGMENTS}

The author thanks R. B. Liu for critical readings of the manuscript.
This work is supported by the National Nature Science Foundation of China, the
Educational Committee of the State Council through the Foundation of Doctoral
Training.


\begin{table}
\caption{
The exponent $\delta$, defined in Eq.(\protect\ref{N^delta}), depends on the
breakdown rule as well as the endurance distribution of the fibers.
}
\begin{tabular}{c|cccccc}
           & $\rho=10$ & $\rho=20$  & $\rho=40$ & $\eta=1$  & $\eta=10$ & $\eta=20$ \\
\tableline
$\beta=1$  & $0.49$    &  $0.83$    & $0.97$    &  $0.17$   & $0.60$    & $0.97$    \\
$\beta=2$  & $0.33$    &  $0.47$    & $0.50$    &  $0.10$   & $0.42$    & $0.52$    \\
$\beta=4$  & $0.23$    &  $0.26$    & $0.27$    &  $0.053$  & $0.23$    & $0.26$    \\
\end{tabular}
\label{tabledelta}
\end{table}

\end{multicols}

\end{document}